# Green Fabrication of Lanthanide doped Hydroxide-based Phosphors: $Y(OH)_3$:$Eu^{3+}$ Nanoparticles for White Light Generation

Tugrul Guner,[a] Anilcan Kus,[a] Mehmet Ozcan,[b] Aziz Genc,[c] Hasan Sahin,[b] and Mustafa M. Demir[a*]

[a] Department of Materials Science and Engineering, Izmir Institute of Technology, Turkey

[b] Department of Photonics, Izmir Institute of Technology, Turkey

[c] Metallurgical and Materials Engineering Department, Faculty of Engineering, Bartin University, 74100 Bartin, Turkey

* Corresponding author: Mustafa M. Demir

T: 00 90 232 750 75 11

e-mail: mdemir@iyte.edu.tr

**Abstract**

Phosphors serve as color conversion layers to generate white light with varying optical features including CRI, CCT, and luminous efficacy. However, they have been produced in harsh synthesis conditions such as high temperature, high pressure, and/or employing a huge amount of solvents. In this work, facile, water-based and a rapid method has been proposed to fabricate lanthanide doped hydroxide-based phosphors. In this sense, submicrometer-sized $Y(OH)_3:Eu^{3+}$ particles, as red phosphor, were synthesized in water at ambient conditions in ≤60 min reaction time. The doping ratio is controlled from 2.5 to 20% in terms of mole. Meanwhile, first principle calculations were also performed on $Y(OH)_3:Eu^{3+}$ to understand the preferable doping scenario and its optoelectronic properties. As an application, these fabricated red phosphors were integrated into PDMS/$YAG:Ce^{3+}$ composite to generate white light. The resulting white light showed a remarkable improvement (≈24%) for LER, a slight reduction of CCT (from 3900K to 3600K), and an unchanged CRI (~60) as the amount of $Y(OH)_3:Eu^{3+}$ increases.

**Introduction**

Lighting is an integral part of civilization and its development has been evolved toward the design of more energy efficient ones with the advancing technology.[1-5] Today, LED-based white light generation becomes already a part of both academic and industrial applications. To obtain white light through LED based configurations, either all three main colors (red-green-blue (RGB)) can be satisfied individually via LED chips or various luminescent materials can be used as color conversion layers over blue or UV LED chips.[6] In this sense, phosphors are commonly used luminescent materials as color conversion layers due to their optical features and high stability.[7-13] Among those, *Cerium doped Yttrium Aluminum Garnet* (YAG:Ce$^{3+}$) is a well-known yellow phosphor that is employed in a LED package together with blue LED to form white light (WLED).[14-19] In such a case, blue light partially absorbed by the phosphor after interacting with the color conversion layer, which is followed by a down-converted yellow emission. Combination of the remaining blue light and yellow emission above the conversion layer generate white light.

WLED system composing of a blue LED and YAG:Ce$^{3+}$ is a facile and cheap way of obtaining white light. However, producing white light through this system shows some inadequate optical properties such as high *Correlated Color Temperature* (CCT) and low *Color Rendering Index* (CRI) due to red deficiency.[20-21] To overcome this problem, one can integrate additional phosphors such as red or red and green over the blue LED or can employ all main RGB colors over UV or near-UV LED chip or can use alternative luminescent materials such as quantum dots,[22] perovskites,[23] organic dyes,[24] etc. In the case of phosphors, combining red phosphor with YAG:Ce$^{3+}$ over blue LED or UV LED is the simplest and cheapest way of increasing CRI while reducing CCT since it saves the YAG-based WLED system from the use of additional phosphors.

Phosphors mostly consist of thermally and chemically stable inorganic hosts such as *YAG*, and rare-earth dopants (*Ce$^{3+}$, Eu$^{3+}$, Dy$^{3+}$, etc.*).[9-11, 13] Visible range emission from these phosphors, such as yellow, green or red, are, in general, is the result of radiative energy transfer between partially filled 4f orbitals of dopant states together with the effective shielding of 5s and 5p orbitals.[25] Moreover, photoluminescence (PL) intensity depends on the transition between 4f -> 4f states; transition from $^5D_0$ to $^7F_1$, $^5D_0$ to $^7F_3$ and $^5D_0$ to $^7F_4$ results in low while transition from $^5D_0$ to $^7F_2$ leading to high PL intensities. Therefore, by manipulating the energy level of the transition states through adjusting the dopant ion, different emission and PL intensities can be obtained.[11, 26-28] These phosphors have been employed in various applications including optoelectronics, field emissive displays and HDTVs, and advanced ceramics.[29-30]

To date, numerous methods for obtaining phosphors to be used either over blue LED or UV LED were reported. However, these materials require harsh synthesis conditions such as high temperature and high pressure, and water-free solvents, which can restrict their commercialization. In this context, several methods have been used widely such as, sol-gel, hydrothermal, combustion, emulsion, precipitation, etc.[31-34] Among those, sol-gel and co-precipitation method are, in general, slow, and usually involve additional steps. On the other hand, there is a huge waste of organic solvents during the emulsion process, which makes this method inefficient in terms of cost and toxicity.[35] Therefore, facile synthesis methods involving water-based reactions at ambient conditions are required in the case of phosphor fabrication. In this study, a promising strategy has been introduced in order to meet these requirements especially in the case of lanthanide doped hydroxide-based phosphor fabrication; it is facile, water-based, and rapid. More specifically, acetate based reagents of both host and dopant are dissolved in LiOH/water solution together under room temperature. Presence of the Li ions during the reaction process may distort the crystal structure and lead to increase in the formation of substitutional defects, which may facilitate the incorporation of dopant ions into the system.[36-39] As an example, luminescent red $Y(OH)_3$:$Eu^{3+}$ phosphors were fabricated via employing this method. Doping process and complete crystallization were achieved in 60 min. Moreover, state-of-the-art first principle calculations were performed on $Y(OH)_3$:$Eu^{3+}$ to investigate its crystallographic structure and resulting electronic and optical properties. In summary, a novel water-based, rapid, and simple method was developed in this study. As an application, red emitting phosphor has been fabricated at ambient conditions in short time. This method can be a promising strategy for fabricating phosphors to be employed as a down-converting materials for pc-converted WLED systems.[33, 40-43]

## 2. Experimental Methodology

*2.1. Materials and Methods*

Yttrium (III) acetate hydrate (Y(Ac)$_3$H$_2$O; >99%), europium (III) acetate hydrate (Eu(Ac)$_3$ H$_2$O; >99%), lithium hydroxide (LiOH; 98%), was purchased from Sigma–Aldrich (St. Louis, MO, USA) and was used as received without any further purification. Cerium doped Yttrium Aluminum Garnet (YAG:$Ce^{3+}$,HB-4155H, Zhuhai HanboTrading Co., Ltd., Guangdong, China) was used as yellow phosphor and PDMS (SYLGARD 184 Kit, Dow Corning, Midland, MI, USA) was used as the polymer matrix. Crystallographic properties of the crystals were enlightened by using X-ray diffractometer (XRD; X'Pert Pro, Philips, Eindhoven, The Netherlands), while their morphology was characterized by Scanning electron microscopy

(SEM; Quanta 250, FEI, Hillsboro, OR, USA). High resolution transmission electron microscopy (HRTEM) micrographs have been obtained by using a FEI Tecnai F20 field emission gun microscope with a 0.19 nm point-to-point resolution at 200 keV equipped with an embedded Quantum Gatan Image Filter (Quantum GIF) for EELS analyses. Images have been analyzed via Gatan Digital Micrograph software. Optical characterization was carried out by using Ocean Optics Spectrometer USB2000+ (Ocean Optics, Duiven, The Netherlands, EU)

*2.2. Synthesis of Eu-doped Yttrium Hydroxide Crystals*

An amount of yttrium (III) acetate hydrate ($5.63 \times 10^{-4}$ moles) and varying amount of europium (III) acetate hydrate were dissolved in 10 mL of deionized water. Depending on the dopant ratio, the amount of europium (III) acetate hydrate was arranged (for instance, to achieve 7.5% $Eu^{3+}$ dopant ratio, $4.55 \times 10^{-5}$ moles of europium (III) acetate hydrate was employed). Subsequently, the mixture was stirred in glass container until appeared transparent. An amount of LiOH (0.04 moles) was added into the transparent solution, respectively. Selecting LiOH as ion source is critical here since the other possible ions such as $Na^+$ and $K^+$ are not as reactive as $Li^+$ ions. Such a high reactivity of $Li^+$ ions in the solution is expected to distort the crystal structure relatively more compared to other possible ions, and therefore can lead to increase in the formation of defects to make doping process more favorable as mentioned already at the end of introduction. The solutions were mixed and sonicated for 5 minutes. After the sonication process, reaction was maintained for 1h at room temperature. Reaction mixture centrifuged twice with water and once with ethanol (5 min, 6000 rpm). After centrifugation, isolated products were dried in oven at 100°C for 1 h.

*2.3. Preparation of PDMS/(Eu-doped Yttrium Hydroxide) Composites*

PDMS (composing of silicon elastomer and curing agent with 10:1 ratio) of 1 g was prepared in a glass vial. Then, fixed amount of 70.0 mg of yellow YAG:Ce$^{3+}$ phosphors was added into the glass vial. In order to investigate the effect of red phosphor amount on the resulting white light properties, desired amount of the red Y(OH)$_3$:Eu$^{3+}$ submicron phosphors in mass was added into the PDMS/YAG:Ce$^{3+}$, and the resulting mixture was stirred until all phosphors mixed homogenously. Then, this mixture dropped into a mold having thickness of 0.2 cm and diameter of 2.0 cm, which are fixed, and the mold was put into vacuum oven for 30 min for solvent evaporation. Finally, it was cured and cross-linked under 100 °C for 30 min, and then the free-standing composite film was removed from the mold.

## 3. Computational Details

To investigate the effect of Europium dopants in the structural and electronic properties of Y(OH)$_3$ crystals, density functional theory-based calculations were also performed using the projector augmented wave (PAW) potentials as implemented in the Vienna ab initio Simulation Package (VASP).[44-47] For the exchange-correlation part of the functional, the generalized gradient approximation (GGA) in the Perdew-Burke-Ernzerhof (PBE) form was employed.[48] In order to obtain the charge transfer between the atoms, the Bader technique was used.[49] The kinetic energy cut-off for plane-wave basis set was taken as 400 eV for all the calculations. For all ionic relaxations, the total energy difference between the sequential steps in the calculations was taken to be $10^{-5}$ eV as the convergence criterion. On each unit cell, the total forces were reduced to a value less than $10^{-4}$ eV/Å. Γ-centered k-point meshes of 2x2x2 were used for 128-atom supercell of bulk Y(OH)$_3$.

**Results and Discussion**

## Structural Characterization of Y(OH)$_3$:Eu$^{3+}$ particles

**Fig. 1a** presents X-ray diffraction pattern of representative Y(OH)$_3$:Eu$^{3+}$ particles having 7.5% dopant ratio (Y(OH)$_3$:(7.5% Eu$^{3+}$)) prepared at various reaction times. The pattern of starting material, i.e. unreacted yttrium acetate is given for the comparison. The reflections can be indexed to hexagonal phase (space group of P63/m) of Y(OH)$_3$ with a lattice constants of a = 6.261 Å, b = 6.261 Å, and c = 3.544 Å satisfying JCPDS: 01-083-2042. The evolution of the crystal is clearly seen in the stack plot of the patterns. When the yttrium acetate is treated with alkaline water for 5 min, its reflections begin to disappear. The extension of synthesis time to 15 min reduces the intensity of the precursor reflections, at the end of 45 min, characteristic signals of the Y(OH)$_3$ labeled with their corresponding planes become more evident. After 60 min reaction time, reflection signals of the resulting product were found to be perfectly matched with the crystallographic data of JCPDS: 01-083-2042. Primitive unit cell of the resulting Y(OH)$_3$ host crystal is presented in **Fig. 1b**. This crystal structure indicates two yttrium atoms located at the one face of the hexagonal phase, and share oxygens saturated with hydrogens. To investigate the crystal formation of the samples fabricated at 5 min, several low magnification HAADF STEM micrographs and a bright field TEM micrograph were taken, and presented in **Fig. 1c.** These micrographs reveal that the sample consists of agglomerated nanoparticles. The clusters formed by the sub-10 nm nanoparticles have sizes between 50 nm to a few microns. It is possible to visualize the individual nanoparticles in the lower left HAADF STEM micrograph where the building blocks of these agglomerates seem to have sizes smaller than 5 nm. Photoluminescence spectrum of the Y(OH)$_3$:(7.5% Eu$^{3+}$) phosphors in varying times was registered at 365 nm excitation wavelength (**Fig. 1d**). At first 5 min, sharp emission signals appear at 592, 595, 613, 616, 690, 697 and 700 nm indicating specific D-D and D-F transitions of Eu$^{3+}$ states may be due to crystal splitting by the Y(OH)$_3$ host. As the synthesis time is extended, the intensity of the signals at 592 and 697 nm shows a significant increase while the ones at 613 and 700 nm are diminishing upon crystallization.

SEM images demonstrate the morphology of the Y(OH)$_3$:(7.5% Eu$^{3+}$) crystals fabricated at varying times in **Fig. 2(a-c)**, from 5 min to 60 min, respectively. The change in morphology of the crystals is evident. The phosphors obtained in 5 min reaction time shows needle-like shape with nano-scale size distribution (**Fig. 2a**). After 15 min of reaction, the crystals grow larger and started to show rod-like structure with sub-micron sizes (**Fig. 2b**). As the reaction time is extended to 60 min, the crystals transform into rice-like structure (**Fig. 2c**). For more detailed information about the morphology of Y(OH)$_3$:(7.5% Eu$^{3+}$) crystals, general TEM and high angle annular dark field (HAADF) STEM micrographs of the sample prepared in 60 min presented in **Fig. 2 (d-f)**. A multipod-like structure (**Fig. 2d**) together with rice-like ones (**Fig. 2 (e-f)**) were obtained. Higher magnifications, as presented in **Fig. 2e**, indicate that multipod-like structures are the result of the reunion of rice-like structures. On the other hand, higher magnification over the rice-like shape demonstrate that these structures have fringed edges (**Fig. 2f**), which may be the ensemble of ~10 nm thickness of individual nanowires.

The level of doping for Y(OH)$_3$:(7.5% Eu$^{3+}$) phosphors produced at 5 min was captured through an ADF STEM micrograph and STEM-EELS analyses of the indicated area presented in **Figure 3**. Elemental composition maps of Y (red) and Eu (green) along with their composites are shown. (Experimental note: The above presented maps are obtained from the Eu M$_{5,4}$ edges located at 1131 eV and 1161 eV and Y L$_{3,2}$ edges located 2080 eV and 2155 eV by using a 1 eV/channel.) The elements of Y and Eu exhibit even distributions throughout the nanoparticle volume with some presence of slightly Eu rich or Y rich regions.

The effect of doping ratio on the photoluminescence properties was investigated. **Fig. 4a** presents the PL spectra of the Y(OH)$_3$:Eu$^{3+}$ red phosphors fabricated at 60 min having various dopant ratios ($\lambda_{Exc}$=365 nm). Characteristic signals of the corresponding transition states of the Eu$^{3+}$ are labeled with A, B, C, D and E to be able to track their changes with respect to adjusted dopant ratios. Initially, at the doping ratio of 2.5%, these signals are comparable with each other in terms of their PL intensity. However, as the doping ratio increases, all emission signals, especially the signals corresponding to A, C, and E show significant increase. The increase of the emission signals of A, C, and E follow paths those grow faster than the remaining B and D. The change of these corresponding emission intensities against doping ratio was presented in **Fig. 4b**. In any case, these paths are saturating with respect to doping ratio, which is expected since the possibility of incorporation of Eu ions into Y(OH)$_3$ has limitation. This limit can be inferred from the **Fig. 4b** as 25-30%. The effect of doping ratio on XRD reflection signals was explored. Since atomic size of Y and Eu is different, the highest reflection at 16° (2$\theta$) corresponding to (100) plane was examined whether any shift is present for the samples of all doping ratios. The crystals that were prepared in 60 min was considered, and their peak positions against their doping ratios was demonstrated in **Fig. 4c**. The corresponding reflections at 16° are fitted with a Gaussian distribution, and then their exact location was registered. As the doping ratio increases, a slight shift of the reflections is observed. This shift may be the result of enlarging lattice of host crystal due to incorporation of Eu$^{3+}$ ions since it has a larger ionic radius of 0.109 nm than the Y$^{3+}$ ions having 0.104 nm.[50-51]

**Tentative Growth mechanism of Y(OH)$_3$:Eu$^{3+}$ Crystals**

Growth mechanism of the resulting Y(OH)$_3$:Eu$^{3+}$ crystals was investigated through considering both SEM and TEM images (Fig. 1c, and Figure 2). The growth mechanism that governs almost the entire nucleation and crystal growth process here was reported already by *Hussain et al.*[54] in detail, where the authors employed hexamethylenetetramine (HMTA) during the fabrication of La(OH)$_3$:Eu$^{3+}$ crystals. In this study, however, HMTA was replaced with LiOH, and Y(OH)$_3$:Eu$^{3+}$ crystals were obtained through the interaction of OH$^-$ ions released from LiOH with Y$^{3+}$ ions from the yttrium source in water at room temperature. Even at 5 min, Y(OH)$_3$:(7.5% Eu$^{3+}$) crystals were being formed as having needle-like shape. Such a fast formation of crystals may hint about the fast nucleation, and here, similar to the idea proposed by *Hussain et al.*[54] (where the authors argued that the excessive NH$_4^+$ ions may accelerate the reaction between La$^{3+}$ and OH$^-$), excessive Li$^+$ ions in the reaction media may be responsible from it. Extending reaction time to 15 min, crystals grow larger and rod-like structure appears in the form of nano-rod bundles probably due to crystal splitting. In the case of 60 min reaction time, rod-like structures transform into rice-like shapes that are in contact mainly with each other as a part of self-assembly process occurring as a result of saturated splitting process.[54] Moreover, multipod-like (or flower-like) morphology was also observed for some particular crystals, which are probably the result of this self-assembly process that finds itself more time to act on these particular crystals. The crystal growth mechanism is summarized and illustrated in **Scheme 1**. Compared to HMTA, which decomposes into ammonia and releases OH$^-$ ions slowly, LiOH is able to give OH$^-$ ions directly to the reaction medium. Therefore, since the rate of hydroxide release is a strong parameter to control size and size distribution and defect content of the resulting crystal, it is expected to obtain variation for these values in the case of comparing the effect of HMTA and LiOH on the crystal growth. Moreover, such a difference between HMTA and LiOH in terms of the rate of OH$^-$ release may allow the formation of these crystals even at room temperature as reported in this study, while the authors that used HMTA were kept the reaction mixture at 75 °C. On the other hand, crystal growth splitting, as observed for various material systems such as SrTiO3,[52] Zn2GeO4,[53] La(OH)3:Eu3+ and La2O3:Eu3+ [54] in literature, is associated with fast crystal growth. A possible cause could be the oversaturation of reactant species. When the concentration of reactive species appears to be higher than a threshold that may vary to each material depending on its chemistry, fast growth of the crystal takes place. Fast growth may force a high density of crystal defects. The atoms do not have enough time for the placement in crystal array and metal atom misplacement may occur during the fast growth. These defects gradually develop nuclei sites developing branches, leading eventually to splitting.

**First principle calculations of the Y(OH)$_3$:Eu$^{3+}$**

Our calculations reveal that ground state structure of Y(OH)$_3$ has hexagonal crystal symmetry with space group P6$_3$/m. **Fig. 5a** demonstrates 14-atom primitive unit cell of Y(OH)$_3$ consists of 2 Yttrium, 6 Oxygen and 6 Hydrogen atoms. Optimized lattice parameters of bulk Y(OH)$_3$ are found to be a = 6.10 Å, b = 6.10 Å and c = 3.51 Å. In this structure each Y atom bonds with nine O atoms with a bond length of 2.39 Å. According to the Pauling scale electronegativity of Y, H and O are 1.22, 2.20 and 3.44, respectively. Bader charge analysis shows that Y(OH)$_3$ crystal structure is formed by 0.73 (0.60) e charge transfer from Y (H) to O atom.

In addition, possible scenarios for Eu doping in Y(OH)$_3$ crystal is also investigated by using state-of-the-art first principles calculations. Total energy minimization calculations suggest that while substitutional doping of Eu atoms by Y, O or H is energetically unfavorable, interstitial doping at the holley site surrounded by the H atoms is a preferable adsorption site as shown in **Fig. 5a**. It is also seen that interstitial doping of Eu atom slightly (0.5%) enlarges the bonds between neighboring atoms belong to Y(OH)$_3$ crystal. Regarding the stability or robustness of Eu dopants in the host material, molecular dynamics calculations show that Eu atoms, covalently attached to the host lattice with binding energy of 2.34 eV, maintain their atomic position in the Y(OH)$_3$ crystal for more than 5 ps at room temperature.

Theoretical calculations show that the electronic structure of the Y(OH)$_3$ host is also significantly modified by the Eu dopant. While the GGA-PBE approximated electronic band dispersion of the host material has a bandgap of 3.83 eV, some midgap states emerge after the interstitial doping of Eu. In addition, the energy bandgap of the host at the vicinity of doped region increases to 4.28 eV. Band and orbital decomposed charge density presented in **Fig. 5b** shows that the midgap state is formed by strongly hybridized Eu and surrounding O atoms. It appears that the strongly bonded Eu atoms not only lead to deformation in the lattice structure but also emergence of defect-like states resulting in additional peaks in the PL spectrum.

**A route for generating white light through integrating red Y(OH)$_3$:Eu$^{3+}$ phosphors into YAG-based color conversion layer**

Red-emitting Y(OH)$_3$:Eu$^{3+}$ submicron phosphors were integrated into YAG-based white LED configuration, and the resulting optical features were investigated. Among varying doping ratios, Y(OH)$_3$:(20% Eu$^{3+}$) fabricated at 60 min was selected as a model since the sample having 20% doping ratio in Fig. 4a provided the highest emission intensities. Optical properties (CRI, CCT, LER, and Lumen) of the PDMS composites are presented in **Figure 6**. **Fig. 6a** shows CRI and CCT of the PDMS composites as a function of the amount of red phosphor. While CRI remains almost unchanged around 60, CCT decreases from 3900 K to 3600 K. The main reason of this inadequate CRI for this system may be the lack of blue color. On the other hand, a significant improvement from 281 to 348 lm/W (nearly 24% increase) is observed for luminous efficiencies (LER) as the red phosphor is employed (**Fig. 6b**). Meanwhile, lumens remains almost fixed showing no any significant change as the red phosphor amount increases. According to color coordinates as presented in **Fig. 6c**, all PDMS composites seem to be accumulated in yellow region. They are far from the white and require blue color to be able to shift resulting color towards white. Therefore, a method based on white LED configuration involving UV LED chip and phosphors was proposed here, which integration of $Y(OH)_3$:(20% $Eu^{3+}$) into PDMS/YAG:$Ce^{3+}$ composite is the part of development towards white light generation. It is possible to obtain remaining main colors, green, and blue, by using the method proposed throughout this study that can be achieved via adjusting the lanthanide. In such a case, by combining all the main colors, one can achieve to generate white light with high CRI and low CCT successfully.

**Conclusion**

A facile, water-based, and rapid synthesis method for the fabrication of lanthanide doped hydroxide-based phosphors at room temperature is presented. Introducing acetate-based metals and lanthanides into the LiOH/water solvent system, which are specifically yttrium and europium acetates for this study, respectively, showed rapid crystallization of lanthanide doped hydroxides. These crystals were driven by crystal splitting growth mechanism that leads to the formation of multipod structure. Even more, this method allows the control of doping ratio by varying dopant concentration. The fabricated red-emitting $Y(OH)_3$:$Eu^{3+}$ phosphors are integrated into well-known YAG-based color conversion layer in order to generate white light. For future applications, this study may lead to fabrication of various phosphors doped with different lanthanides that can result in a targeted color. In this sense, one can produce white LED system composing of UV-LED chip with red-green-blue phosphors since the method presented here offers easy, cheap and environmental-friendly synthesis for the fabrication of these phosphors.

**Corresponding Authors**

* Prof. Mustafa M. DEMIR. Material Science and Engineering Department, Izmir Institute of Technology, Izmir, Turkey.

**Author Contributions**

The manuscript was written through contributions of all authors. All authors have given approval to the final version of the manuscript.

**Acknowledgements**

**FIGURE CAPTIONS**

**Scheme 1.** Schematic demonstration of the crystal growth mechanism for the development of multipod-like structure of the Y(OH)$_3$:Eu$^{3+}$ phosphors.

**Figure 1. (a)** XRD patterns of Y(OH)$_3$:(7.5% Eu$^{3+}$) phosphors prepared at various reaction times; 5, 15, 45, and 60 min, **(b)** schematic presentation of the unit-cell of Y(OH)$_3$ lattice, **(c)** general HAADF STEM and TEM micrographs of the Y(OH)$_3$:(7.5% Eu$^{3+}$) particles prepared in 5 min showing the presence agglomerated nanocrystals, and **(d)** PL spectrum of Y(OH)$_3$:(7.5% Eu$^{3+}$) phosphors.

**Figure 2.** SEM images of the Y(OH)$_3$:(7.5% Eu$^{3+}$) phosphors obtained at different synthesis times; **(a)** 5 min, **(b)** 15 min, and **(c)** 60 min. **(d-f)** present the overview TEM and HAADF STEM micrographs of the particles prepared in 60 min. The sample is composed of micron-sized multipods, which seem to be controlled ensemble of ~10 nm thick nanowires.

**Figure 3.** Annular dark field (ADF) STEM micrograph of an agglomerate of nanoparticles synthesized at 5 min with 7.5% doping ratio. STEM-EELS elemental composition maps of the area indicated with a white rectangle: Y (in red) and Eu (in green) maps along with their composite image.

**Figure 4. (a)** PL spectrum of the Y(OH)$_3$:Eu$^{3+}$ phosphors synthesized for 60 min with different doping ratios, **(b)** variation of their corresponding emission peak intensities, labeled with A, B, C, D, and E, with respect to doping ratio, and **(c)** the shift of the $2\theta$ reflection position of 16° with respect to doping ratio of the particles.

**Figure 5.** **(a)** Perspective view for the atomic structure of Europium doped $Y(OH)_3$, **(b)** (left) the electronic band dispersion of Eu doped $Y(OH)_3$ (The Fermi level is set to zero.), and (right) top and side view for charge density corresponds to the midgap electronic state.

**Figure 6.** WLED application of different amount of red $Y(OH)_3$:$Eu^{3+}$ phosphors fabricated at 60 min having 20% doping ratio combined with YAG:$Ce^{3+}$ phosphor and their corresponding **(a)** CRI, CCT, **(b)** LER, Lumens, and **(c)** CIE color coordinates.

# FIGURES and TABLES

# Scheme 1

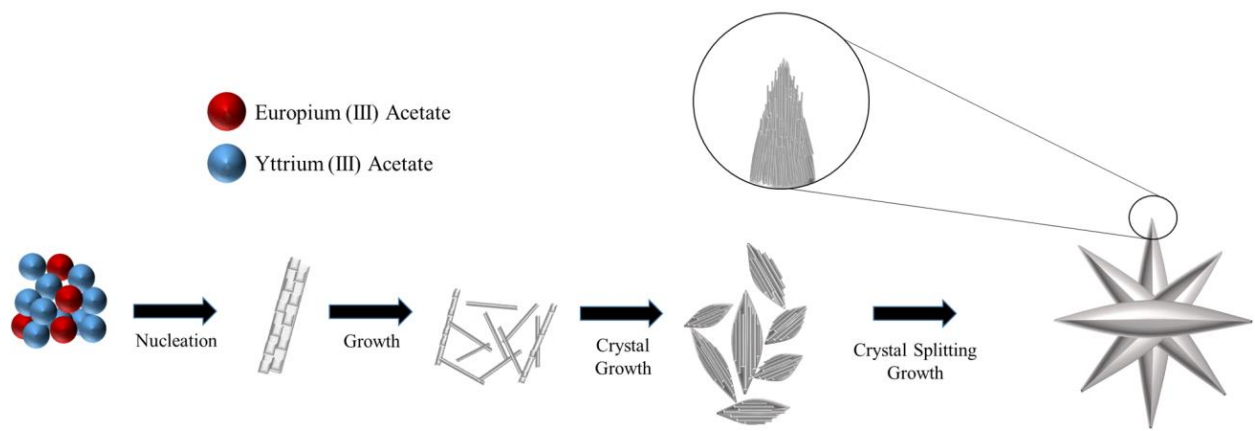

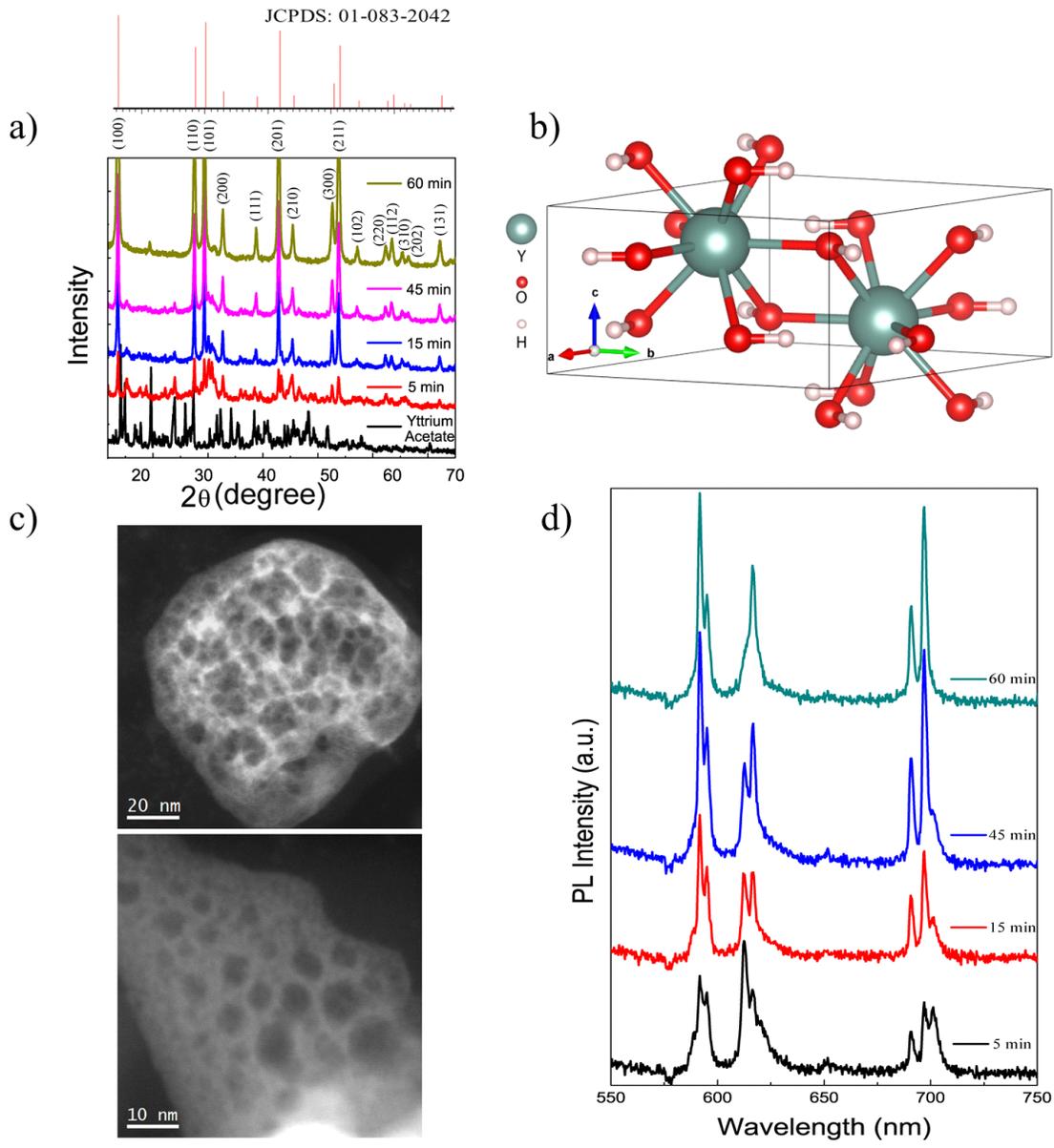

**Figure 2**

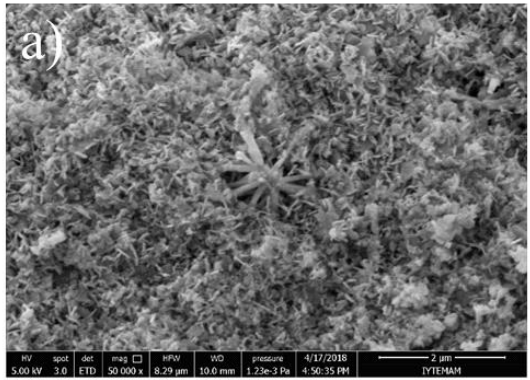
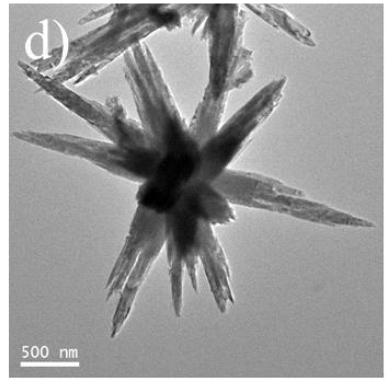
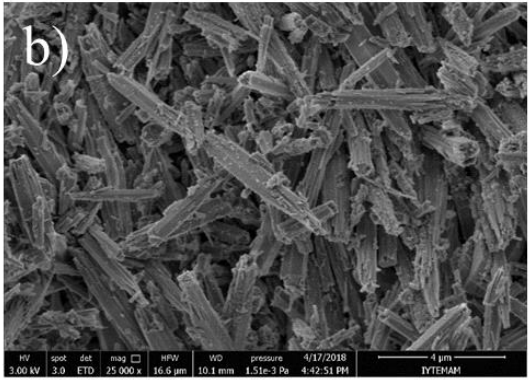
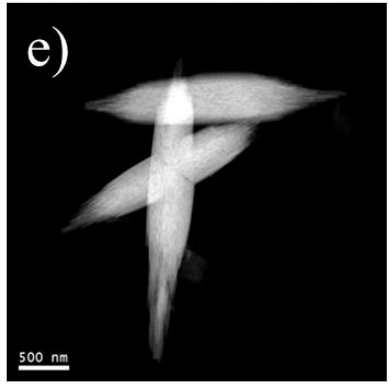
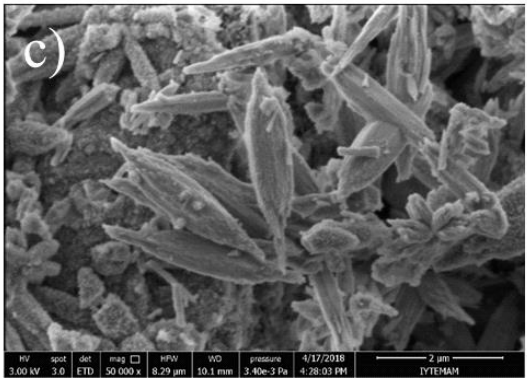
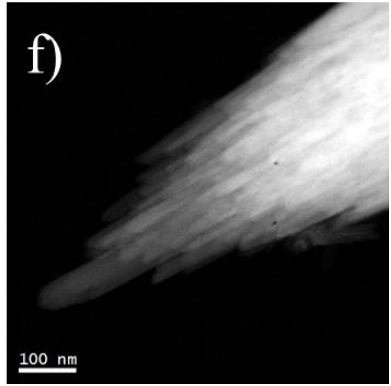

**Figure 3**

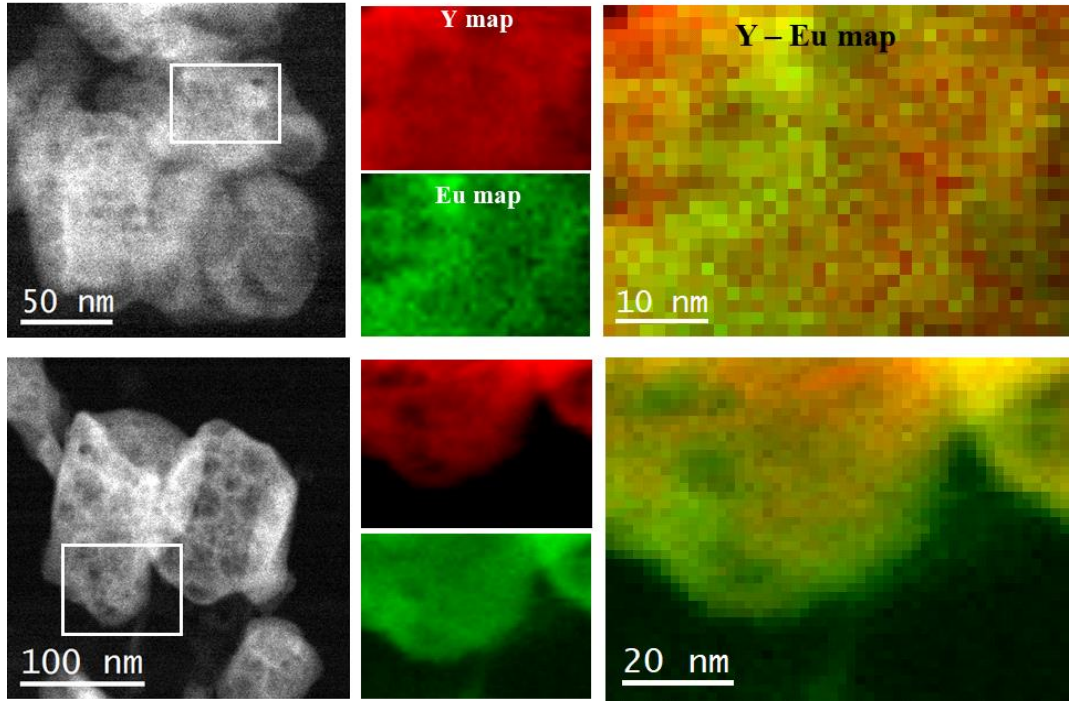

**Figure 4**

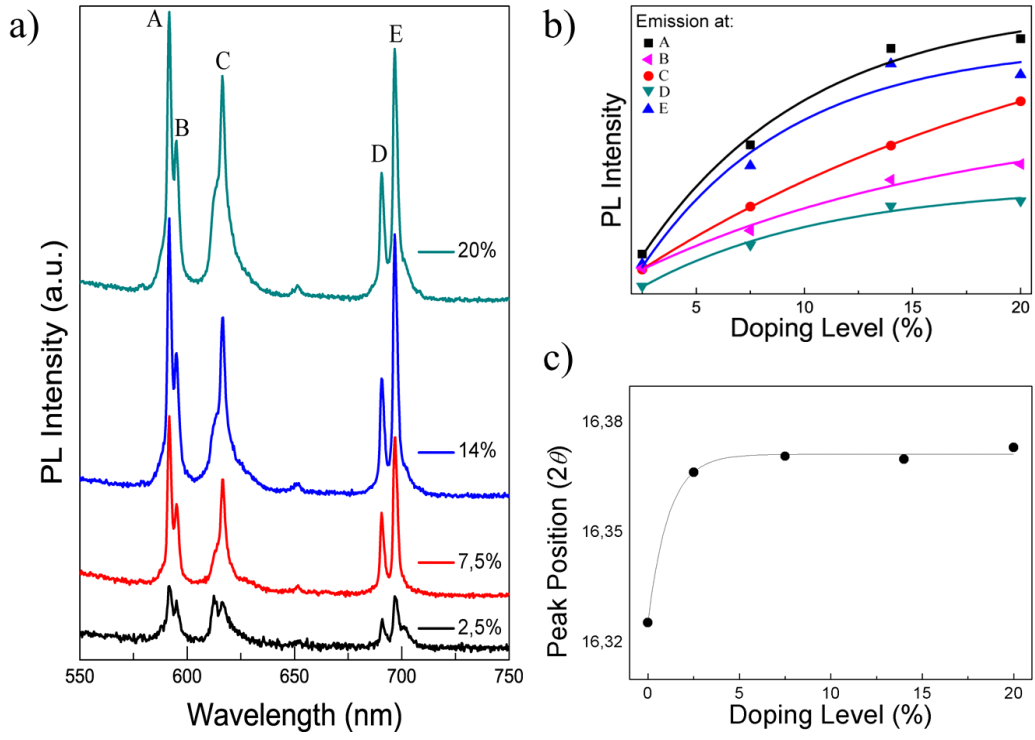

**Figure 5**

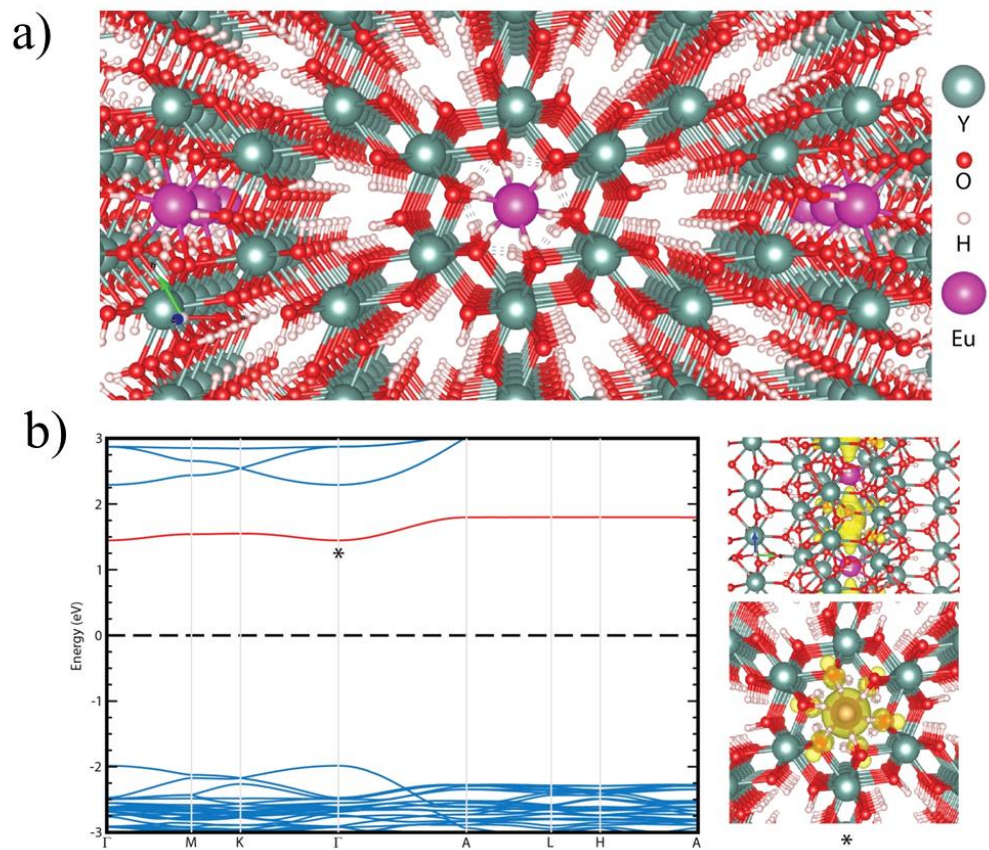

**Figure 6**

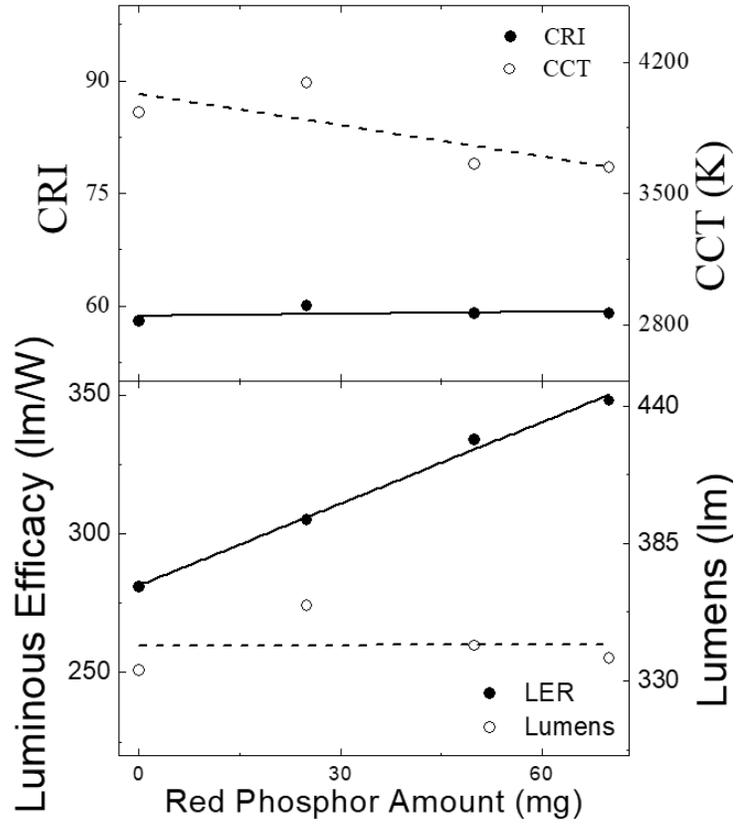
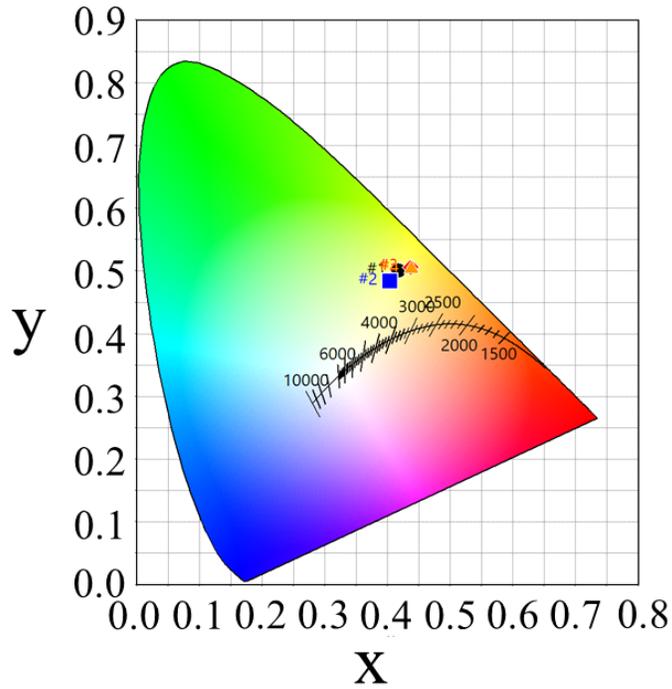